\documentclass[amsmath,amssymb,aps,prl,reprint,groupedaddress]{revtex4-1}
\usepackage {CJK}
\usepackage[utf8]{inputenc}
\usepackage{graphicx}
\usepackage{dcolumn}
\usepackage{epstopdf}
\usepackage{pgfplots}
\pgfplotsset{plot coordinates/math parser=false}
\newlength\figureheight
\newlength\figurewidth
\pgfplotsset{compat=newest}
\usetikzlibrary{plotmarks}
\usetikzlibrary{arrows.meta}
\tikzset{>=latex}

\usepackage{grffile}
\usepackage{amsmath}
\usepackage{standalone}

\usepackage[breaklinks=true]{hyperref}
\usepackage{breakcites}

\begin{document}
\begin {CJK*} {GB} { }

\title{Proof of concept for an optogalvanic gas sensor for NO based on Rydberg excitations}


\author{J. Schmidt$^{1,2}$, M. Fiedler$^1$, R. Albrecht$^1$, D. Djekic$^3$, P. Schalberger$^2$, H. Baur$^2$, R. L\"ow$^1$, N. Fruehauf$^2$, T. Pfau$^1$, J. Anders$^3$, E. R. Grant$^4$ and H. K\"ubler$^1$}
\email[]{h.kuebler@physik.uni-stuttgart.de}
\homepage[]{www.iqst.org}
\affiliation{$^1$5. Physikalisches Institut and Center for Integrated Quantum Science and Technology (IQST), Universit\"at Stuttgart, Pfaffenwaldring 57, 70569 Stuttgart, Germany\\
$^2$Institut f\"ur Gro{\ss}fl\"achige Mikroelektronik and IQST, Universit\"at Stuttgart, Allmandring 3b, 70569 Stuttgart, Germany\\
$^3$Institute of Smart Sensors and IQST, Universit\"at Stuttgart, Pfaffenwaldring 47, 70569 Stuttgart, Germany\\
$^4$Department of Chemistry and Department of Physics \& Astronomy, The University of British Columbia, 2036 Main Mall, Vancouver, BC Canada V6T 1Z1}


\date{\today}

\begin{abstract}
 We demonstrate the applicability of 2-photon Rydberg excitations of nitric oxide (NO) at room temperature in a gas mixture with helium (He) as an optogalvanic gas sensor. The charges created initially from 
 succeeding collisions of excited NO Rydberg molecules with free electrons are measured as a current on metallic electrodes inside a glass cell and amplified using a custom-designed high-bandwidth transimpedance amplifier attached to the cell. We find that this gas sensing method is capable of detecting NO concentrations lower than {10 ppm} even at atmospheric pressures, currently only limited by the way we prepare the gas dilutions.
\end{abstract}
\maketitle
\end{CJK*}
\section{Introduction}
Nitric oxide (NO) plays an important role in a variety of biological and chemical processes. Since 1980 
 \cite{Furchgott1980, Ignarro1987, Moncade2006} it has been known, that NO is responsible for the vasodilation, i.e. the relaxation of arterial smooth muscles, and since then 
 its vital role for the immune system and as neurotransmitter has created increased attention. More specifically, NO can move freely between cell membranes to stimulate RNA and protein synthesis, facilitate neurotransmission and control gene expression. Because NO is a radical it is supporting the immune system by destroying e.g. foreign bacteria and more recent work has shown that it regulates the immune function of macrophages \cite{Yun1996}. Yet, in excessive amounts NO is neurotoxic and can hasten apoptosis \cite{Thomas2008}. Consequently, the detection of NO is of great interest for the diagnosis of inflammatory diseases such as asthma, where NO is detected in the exhaled air using chemiluminescence \cite{Alving1993, Jorissen2001}. Moreover, NO can serve as a signaling molecule in carcinogenesis and tumor growth, e.g. breast cancer \cite{Choudhari2012, HAKLAR2001}, where it is most often detected by amperometric measurements using a platinum electrode \cite{Jensen2013}. However, the drawback of the latter approach is the dependence on the pH level of the surrounding tissue and large cross-sensitivities to other small molecules such as CO. 
 Compared to the chemiluminescence method, the extraction step of NO in the gaseous phase can be omitted, which greatly simplifies the experimental setup.
 Although the chemiluminescence method is the most sensitive detection scheme for NO, it can only be competitive for total gas volumes exceeding ${1 \text{ l}}$ \cite{Bates1992}. Other gas sensing methods such as purely optical sensors based on quantum cascade lasers in the mid-infrared regime \cite{Menzel2001} or e.g. field-effect transistors using gold nanoparticles as catalytically active material \cite{DiFranco2009}, are either prone to fluctuations in the light level, show cross-sensitivities, are easily destroyed in chemically harsh environments or show only a low dynamic range and have altogether no competitive sensitivity.
 
As an approach to overcome most of the aforementioned problems of the currently used methods, we propose and
 experimentally demonstrate the use of a gas sensor based on Rydberg excitations, which combines the advantages of optical and amperometric methods. In our approach, we optically excite Rydberg states in thermal NO, which are close to the ionization continuum but where the electron is still bound. The Rydberg molecules decay mainly due to collisions with free electrons into a pair of charges \cite{Saquet2012, Haenel, Sadeghi, PCCP, Morrison}. Those charges are then measured as a current using two electrodes. 
Since we are using several laser transitions to excite the molecules into the Rydberg state, the method is 
 extremely selective for the molecule of interest. Therefore, the occurrence of a current is a clear and unique indication of the presence of NO. Moreover, the sensor is immune against light fluctuations, because the populations of the levels involved are saturated by using sufficiently intense light fields.
 Furthermore, we are able to perform the measurement in a single shot in a volume as small as $10^{-4}\text{ l}$ for the presented prototype. In the first proof of principle experiments we achieve a detection limit as low as {10 ppm} only limited by the way we prepare the gas dilutions. Additionally, an applicability of this method even at atmospheric pressures is demonstrated.
\section{Experimental setup}
A pair of Nd:YAG pumped dye lasers produce a double-resonant sequence of nanosecond short pulses, $\omega_1+\omega_2$ to form a state-selected 
 Rydberg gas of NO. The first pulse, with a wavelength of 225 nm, populates lower-lying rotational levels of the 
 $\text{A } ^2\Sigma$ $(v=0)$ intermediate state in transitions from a room-temperature distribution of $\text{X } ^2\Pi_{1/2}$ ground state NO. Set to a pulse energy of $(4.5\pm0.3)\text{ }\mu\text{J}$ and a $1/e^2-$radius of {$(1.7\pm0.3)$ mm}, $\omega_1$ has sufficient power to approach saturation of the X to A transition while avoiding indiscriminate two-photon ionization of NO or other possible gases, driven by $\omega_1$ alone. With a tuneable wavelength around 328 nm, the second laser excites A-state molecules to a dense manifold of Rydberg levels within an interval 140 cm$^{-1}$ below the $\text{A } ^1\Sigma$ $(v^+=0)$ ground state of the NO$^+$ ion \cite{Miescher, Jungen, Jungen_ryd}.  Many of the Rydberg states populated by tuning $\omega_2$ predissociate to neutral N and O atoms. Only those in the series $nf(N^+)$ live long enough to produce an ionization current. Here, $N^{+}$ represents an ion rotational state convergence limit determined by the A-state rotational distribution prepared in the first step of double resonance. With a pulse energy of $(10.0\pm0.5)\text{ mJ}$ and a beam radius of $(1.5\pm0.1)\text{ mm}$, $\omega_2$ saturates the A-Rydberg transitions.
\begin{figure}[htbp]
	\hspace{-13.5pt}
	\begin{tikzpicture}
	\node[anchor=south west] (myplot) at (0.55,0.8) {
		\input{Rydberg.tex}
	};
	\begin{scope}[x={(myplot.south east)}, y={(myplot.north west)}]
	%
	%
	%
	\draw[<-,thick,black] (.4,.86)--(.45,.88) 
	node[right,align=center,text=black] 
	{n=40};
	\draw[thick,black] (.04,.30) 
	node[right,align=center,text=black,rotate=90] 
	{Ions per shot $Q_{ion}$ [$10^8$]};
	\draw[thick,black] (0.35,.06) 
	node[right,align=center,text=black] 
	{Wavelength $\omega_2$ [nm]};
	\draw[thick,black] (0.672,.85) 
	node[right,align=left,text=black] 
	{$2.5\times10^{-4}$};
	\draw[thick,black] (0.685,.65) 
	node[right,align=left,text=black] 
	{$6.2\times10^{-5}$};
	\draw[thick,black] (0.7,.45) 
	node[right,align=left,text=black] 
	{$6.2\times10^{-6}$};
	\draw[thick,black] (0.71,.25) 
	node[right,align=left,text=black] 
	{Evacuated};
	%
	%
	\end{scope}
	\node[anchor=south west, inner sep=0pt] (schematic) at (0.,6.5)
	{\includegraphics[width=0.95\columnwidth]{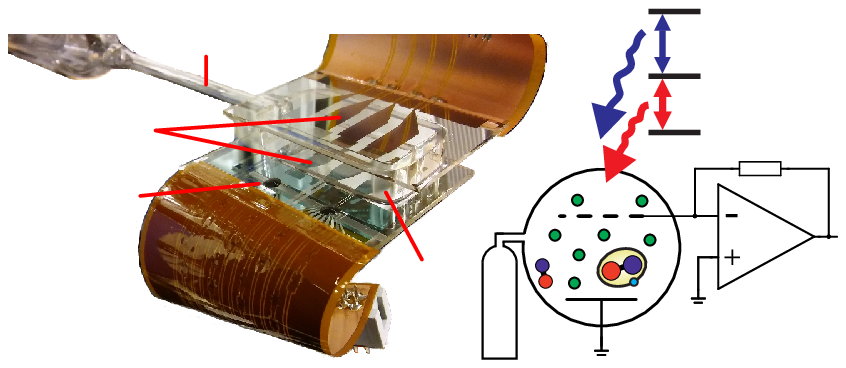}};
	\begin{scope}[shift = {(schematic.south west)}, x={(schematic.south east)}, y={(schematic.north west)}]
	%
	%
	%
	\draw[thick,black] (.0,.95)
	node[right,align=center,text=black] 
	{(a)};			
	\draw[thick,black] (.65,.95)
	node[right,align=center,text=black] 
	{(b)};			
	\draw[thick,black] (.01,-.04)
	node[right,align=center,text=black] 
	{(c)};
	\draw[thick,black] (.15,.88)
	node[right,align=center,text=black] 
	{\footnotesize{Gas inlet}};			
	\draw[thick,black] (-.01,.65)
	node[right,align=center,text=black] 
	{\footnotesize{Electrodes}};			
	\draw[thick,black] (.435,.2)
	node[right,align=center,text=black] 
	{\footnotesize{Quartz}\\\footnotesize{frame}};
	\draw[thick,black] (-0.01,.45)
	node[right,align=center,text=black] 
	{\footnotesize{Amplifier}};
	\draw[thick,black] (.82,.97)
	node[right,align=center,text=black] 
	{\footnotesize{NO$^+$ X $^1\Sigma^+$}};		
	\draw[thick,black] (.82,.78)
	node[right,align=center,text=black] 
	{\footnotesize{A $^2\Sigma^+$}};
	\draw[thick,black] (.82,.625)
	node[right,align=center,text=black] 
	{\footnotesize{X $^2\Pi_{1/2}$}};
	\draw[thick,black] (.78,.88)
	node[right,align=center,text=black] 
	{\footnotesize{$\omega_2$}};	
	\draw[thick,black] (.78,.7)
	node[right,align=center,text=black] 
	{\footnotesize{$\omega_1$}};
	\draw[thick,black] (.591,-.02)
	node[right,align=center,text=black,rotate=90] 
	{\footnotesize{$\text{He}/\text{NO}$}};
	\draw[thick,black] (.92,.25)
	node[right,align=center,text=black] 
	{\footnotesize{$U_{out}$}};
	\end{scope}
	\end{tikzpicture}
	\caption{
		\label{RydbergSeries}
		(a) Picture of the glass cell.
		(b) Schematic of the experimental setup. The glass cell is being filled with a mixture of NO and He, which is prepared and stored in a stainless steel vessel. NO is excited to a Rydberg state via the two laser pulses $\omega_1$ and $\omega_2$. The charges emerging from the ionization of the excited molecules are detected and amplified with a TIA.
		(c) Rydberg series from $n=35$ to $n=44$ of thermal NO in a buffer gas of helium at a pressure of 11 mbar. From top to bottom the ratio of NO molecules to He atoms decreases from $2.5\times10^{-4}$ down to a spectrum recorded in a presumably evacuated cell.}
\end{figure}

The excited Rydberg molecules decay into charges mainly via 
 subsequent collisions with free electrons, which is different from groundstate collisional ionization of Rydberg states in e.g. alkali atoms \cite{Barredo2013}. We collect the emerging charges using two metallic electrodes \cite{Barredo2013, Seaver1983, Ebata1983} and convert this current to a voltage using a custom-designed transimpedance amplifier (TIA) \cite{Anders2017}.
The advantage of exciting a transition, which is only close to the ionization continuum and ionizes only subsequently is that we
 gain a much higher selectivity compared to a direct photoionization. As a result no further means to distinguish between e.g. NO and NO$_2$ have to be taken \cite{McKeachie2001, Short2006}.
We shine the lasers in a co-propagating manner into the cell through a transparent glass frame made from quartz glass in order to
 allow transmission of the $\omega_1$ laser light at {225 nm}. The volume inside the frame has a length of {$(15.0\pm0.2)$ mm}, which yields an excitation volume of $(100.4\pm17.9) \text{ mm}^3$. A quartz tubing with a {KF-16} vacuum flange 
 is attached to the frame, through which the gas mixtures are filled into the cell. Above and below the frame are two glass substrates glued with {Epotek-301} on which the electrodes consisting of {150 nm} chromium are realized in standard thin-film technology. 
Since NO is not as reactive as e.g. alkali metals which are more commonly used for experiments in Rydberg physics, no special  sealing method had to be applied here \cite{Daschner2012, Daschner2014, Schmidt2017}.
Onto one of the substrates, outside the future cell part, we deposited the footprint for the TIA as well as the connections for
 the flat band cable leading to the voltage supplies and the scope for readout of the current-to-voltage converted signal $U_{\text{out}}$. Both the TIA and the flat band cable were bonded to the glass substrate using an anisotropic conductive film. The TIA is encapsulated with black optics epoxy, which helps to prevent leakage currents caused by humidity on the glass surface as well as to reduce the cross-sensitivity to stray light. A picture of the cell is shown in Fig. \ref{RydbergSeries} a).
The working principle of the TIA was presented in \cite{Anders2017}. 
 The custom TIA, which is based on a tunable pseudoresistor features an on-chip compensation capacitor and a tunable transimpedance between {120 dB$\Omega$} and {180 dB$\Omega$}, which is specifically designed to be immune against temperature variations.
 The tunable gain allows for an extremely large dynamic range of the gas sensor. For the present realization of the gas sensor we set the transimpedance gain to $126 \text{ dB}\Omega$, which permits a bandwidth of {2 MHz} and an input referred current noise of at most $10^{-12} \text{ A}/\sqrt{\text{Hz}}$ at {1 MHz}. 
 The high bandwidth of the amplifier is required to preserve the temporal features in the measured current produced by the pulsed laser excitation, which allows a measurement of the arrival times of the ionic charges.
 The input of the amplifier is biased at a potential of {1.25 V} with respect to the counterelectrode, such that an NO$^+$ ion created in the middle of the cell in a distance of {2.5 mm} from the electrodes would arrive after a flight time of roughly {900 ns} in vacuum. This corresponds to an expected maximum frequency component in the current pulse of {1.1 MHz}. This frequency will be somewhat lower because of the smaller mean free path length of the ions in a buffer gas. The arrival time of an electron will be 230 times faster than the ionic signal and can therefore not be observed.
For each laser shot we record a voltage trace, examples of which are shown in {Fig. \ref{SignalShapes}}. With the known 
 amplification factor we calculate the amount of ionic charges $Q_{\text{ion}}$ per shot by integrating over the whole voltage trace. By scanning the wavelength of the $\omega_2$ laser we can record spectra of the Rydberg series of thermal NO in a buffer gas of He like the ones shown in {Fig. \ref{RydbergSeries} c)}. Since the pulsed dye lasers are not seeded by a narrow linewidth laser, they are not Fourier limited and  can easily exceed a linewidth of {100 GHz}. Therefore, also intermediate states lying nearby the 
 $\text{A }^2\Sigma\leftarrow\text{X }^2\Pi_{1/2}$ transition in the $Q$-branch, such as the ones with rotational quantum number $N'=1...5\leftarrow N=1...5$ are populated and converge to the {$nf(N^+)$} Rydberg series with their own quantum defect. This leads to all the smaller peaks besides the main 
 $\text{X }^1\Sigma^+ \leftarrow\text{A }^2\Sigma\leftarrow\text{X }^2\Pi_{1/2}$ transition.
 
Different dilutions are prepared by filling a small volume with a certain pressure of NO and letting this expand into a
 much larger volume, which is then filled up to a certain pressure with He. That means, we are changing the amount of NO molecules while keeping the amount of He atoms constant compared to the NO amount. This dilution is stored in a stainless steel vessel and a small portion of it is used for the experiments at an absolute pressure of $(11\pm3) \text{ mbar}$. Between each dilution preparation the whole apparatus is evacuated using a rotary vane pump. 

\section{Results}
\begin{figure}[htbp]
	\centering
	\input{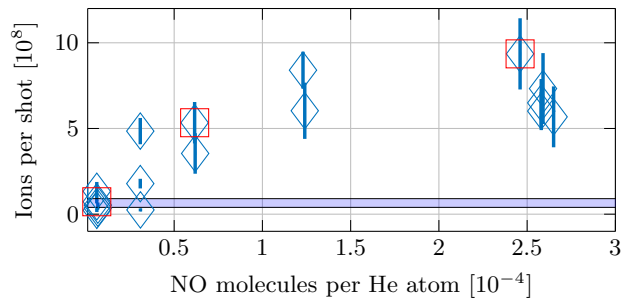}
	\caption{
		\label{ChargeVsDilutions}
		Number of ions at each laser shot as a function of the ratio between NO molecules and He atoms at a constant pressure of {11 mbar}. The horizontal blue beam represents the number of charges measured for an evacuated cell. The corresponding spectra of the red marked data points are shown in Fig. \ref{RydbergSeries} (c).}
\end{figure}

The maximum number of ions per shot obtained for a Rydberg state with main quantum number ${n=40}$ is plotted in 
 Fig. \ref{ChargeVsDilutions} for a constant pressure of {11 mbar}. For each dilution the measurements were taken on a identically prepared sample. The large fluctuations in the measurement data are partly caused by difficulties in the sample preparation, which we discuss below. Another share can be attributed to the insufficient frequency stability of the $\omega_1$ laser.

\begin{figure}[htbp]
	\centering
	\input{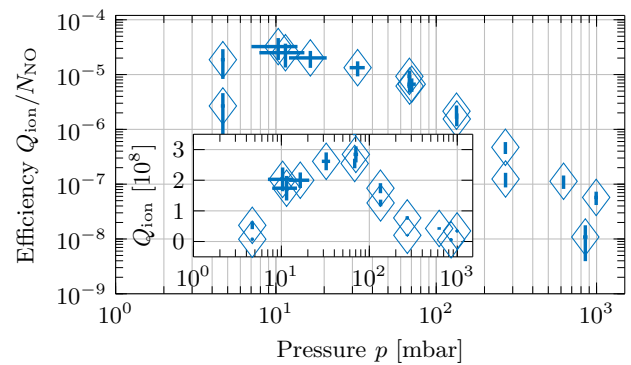}
	\caption{
		\label{EfficiencyVsPressure}
		Efficiency represented as the ratio of the number of detected charges and the number of nitric oxide molecules, which were presumably present in the excitation volume as a function of the overall pressure for a single dilution consisting of $2.0\times10^{-5}$ NO molecules per He atom. The inset shows the same data values, but translated to the number of ions detected in a single shot as function of the pressure.
	}
\end{figure}

Following this series of measurements at constant pressure we performed another series to measure the pressure dependency. We prepared a mixture consisting of $2.5\times10^{-4}$ NO molecule per He atom. The number of ions measured per shot as a function of the the overall pressure inside the glass cell is depicted in the inset of Fig. \ref{EfficiencyVsPressure} again for $n=40$.
 
We observe that the total number of detected charges at a pressure of {11 mbar} is lower than we recorded under the same conditions for the dilution measurement shown in Fig. \ref{ChargeVsDilutions}.
We attribute this to the adsorption of NO molecules onto the surface of the metal vessel used for storing the dilution during the measurement. 
NO forms a nitrosyl complex with transition metals \cite{McCleverty1979, Eisenberg1975, Ford2002}. In this complex NO is only weakly bound and, therefore, NO molecules diffuse into the metal \cite{SCHMICK1982471}. The direction of this diffusion depends on the concentration of NO inside the metal and in the gas phase \cite{PONTRELLI20073658}. At low enough pressures a metal surface, which was once contaminated with NO can outgas NO as well. As a consequence, this outgassing process deteriorates any dilution which is thinner than approximately {10 ppm} and sets a technical detection limit to this demonstration of our NO detection scheme. This limit is marked in Fig. \ref{ChargeVsDilutions} as a blue beam representing the number of charges measured for a cell, which was supposedly pumped empty.
 
\begin{figure}[htbp]
 	\centering
 	\input{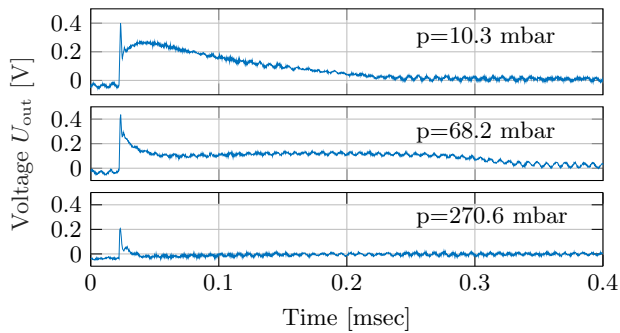}
 	\caption{
 		\label{SignalShapes}
 		Voltage $U_{out}$ at the output of the TIA as a function of time for different pressures of the same dilution.
 	}
 \end{figure}

In order to give an upper estimation of the excitation efficiency $Q_{\text{ion}}/N_{\text{NO}}$ we scale the number of detected ions with the amount of supposedly prepared NO molecules. The excitation efficiency as a function of the absolute pressure $p$ is plotted in the main part of Fig. \ref{EfficiencyVsPressure}. In combination with Fig. \ref{ChargeVsDilutions} such a graph can be used as a worst case estimation of the amount of expected charges for a certain dilution at a specific pressure. Across a pressure range from {4 mbar} to {1000 mbar} the excitation efficiency changes over 2.5 orders of magnitude.
The measured excitation efficiency increases first due to increasing Rydberg-Rydberg collision events and decreases again for higher
 pressures. This is a consequence of collisional deexcitation of the intermediate state but also due to recombination of the created charges during their flight time to the electrodes. This can be verified by looking at the single voltage traces $U_{out}$ after each laser shot in Fig. \ref{SignalShapes} and comparing them for different pressures. 

The signal rise time is on the order of $4 \text{ }\mu\text{s}$, which is well covered by the bandwidth of the TIA. The maximum
 amplitude of the signal decreases for higher pressures because of the lower excitation efficiency. Furthermore, the tail of the signal shifts to longer times as a result of the decreasing mean free path length of the ions inside the background gas. This allows for more recombination processes and therefore a loss of detectable charges.
 
The maximally achievable excitation efficiency can be estimated if one assumes that the groundstate 
$\text{X }^2\Pi_{1/2}$ is thermally populated with a fraction ${\frac{hcB}{kT}(2J+1)e^{-BJ(J+1)hc/kT}}$ of all NO molecules \cite{herzberg1945}, neglecting the minor contribution of the thermal distribution of the vibrational states at $T=300\text{ K}$ and with the rotational constant $B=1.671854\text{ cm}^{-1}$ \cite{Hall1966}. Since there is no coherence between the groundstate and the intermediate state, only one fourth of the atoms in the groundstate can be excited to the Rydberg state. Therefore the maximally achievable efficiency is $4.0\times10^{-3}$, as long as the $\omega_1$ laser populates only one single intermediate state, the collisional ionization probability stays one and there are no recombination processes occurring. 

\section{Summary}
We demonstrated how selective Rydberg excitations can potentially be used as an optogalvanic gas sensor for a molecule of high biological
relevance. The determined detection limit of {10 ppm} is not limited by the excitation efficiency or the detection of the charges but rather by the way we prepared the NO-He dilutions. For the future we will therefore construct a mixing apparatus which works in through-flow rather than in a static mode and where stainless steel parts are omitted as much as possible. This will allow for saturating all surfaces with NO and therefore having merely a constant offset on top of the charge measurements.
Another important issue is the large linewidth of the pulsed dye lasers, which spoils the selectivity of the proposed gas sensor. 
This can be regained by employing state of the art cw lasers. A cw excitation would also relax the requirements on the bandwidth of the amplification circuitry, 
 which in turn would allow for an increased TIA gain and therefore lower noise; thereby improving the detection sensitivity of the readout electronics.
For a practical application such as breath analysis, a volume flow of exhaled breath of \mbox{50 ml/second} is recommended \cite{ATS2005}. With the measured 
 Ryd\-berg excitation efficiency of approximately $10^{-7}$ at a pressure of {1 bar} and a minimum detectable current of {100 fA} with the employed amplifier, which is far above the noise floor, a detection limit of {5 ppb} could in principle be achieved at the flow rate mentioned above.
Although the measured sensitivities in this proof of principle experiment are still orders of magnitude worse compared to other 
 gas sensors such as chemiluminescence based sensors, the presented results indicate the potential of the proposed Rydberg detection scheme to provide a promising sensitivity and a very good selectivity at atmospheric pressures and in chemically harsh environments.

\section{Acknowledgements}
This work was supported by the US Air Force Office of Scientific Research (Grant No. FA9550-17-1-0343), together with the Natural Sciences and Engineering research Council of Canada (NSERC), the Canada Foundation for Innovation (CFI), the British Columbia Knowledge Development Fund (BCKDF) and the German Research Foundation (DFG SPP 1929 GiRyd).

\bibliography{literatur}

\end{document}